\documentclass[acmtog]{acmart}

\AtBeginDocument{%
  \providecommand\BibTeX{{%
    \normalfont B\kern-0.5em{\scshape i\kern-0.25em b}\kern-0.8em\TeX}}}

\begin{document}
\title{CNN-Assisted Steganography - Integrating Machine
Learning with Established Steganographic Techniques}

\author{Andrew Havard}
\email{ahavard@smu.edu}
\affiliation{%
  \institution{Darwin Deason Institute for Cyber Security, Southern Methodist University}
  \city{Dallas}
  \state{Texas}
  \country{USA}
  \postcode{75275-0001}
}

\author{Theodore Manikas}
\email{manikas@lyle.smu.edu}
\affiliation{%
  \institution{Darwin Deason Institute for Cyber Security, Southern Methodist University}
  \city{Dallas}
  \state{Texas}
  \country{USA}
  \postcode{75275-0001}
 }

\author{Eric C. Larson}
\email{eclarson@lyle.smu.edu}
\affiliation{%
  \institution{Darwin Deason Institute for Cyber Security, Southern Methodist University}
  \city{Dallas}
  \state{Texas}
  \country{USA}
  \postcode{75275-0001}
}

\author{Mitchell A. Thornton}
\email{mitch@lyle.smu.edu}
\affiliation{%
  \institution{Darwin Deason Institute for Cyber Security, Southern Methodist University}
  \city{Dallas}
  \state{Texas}
  \country{USA}
  \postcode{75275-0001}
}

\renewcommand{\shortauthors}{Havard, et al.}

\begin{abstract}
We propose a method to improve steganography by increasing the resilience of stego-media to discovery through steganalysis. Our approach enhances a class of steganographic approaches through the inclusion of a steganographic assistant  convolutional neural network (SA-CNN).  Previous research showed success in discovering the presence of hidden information within stego-images using trained neural networks as steganalyzers that are applied to stego-images. Our results show that such steganalyzers are less effective when SA-CNN is employed during the generation of a stego-image. We also explore the advantages and disadvantages of representing all the possible outputs of our SA-CNN within a smaller, discrete space, rather than a continuous space. Our SA-CNN enables certain classes of parametric steganographic algorithms to be customized based on characteristics of the cover media in which information is to be embedded.  Thus, SA-CNN is adaptive in the sense that it enables the core steganographic algorithm to be especially configured for each particular instance of cover media. Experimental results are provided that employ a recent steganographic technique, S-UNIWARD, both with and without the use of SA-CNN.  We then apply both sets of stego-images, those produced with and without SA-CNN, to an exmaple steganalyzer, Yedroudj-Net, and we compare the results. We believe that this approach for the integration of neural networks with hand-crafted algorithms increases the reliability and adaptability of steganographic algorithms.
\end{abstract}




\maketitle

\section{Introduction}
Steganography is the process of hiding information within some medium such as an image or audio file.  A converse procedure, steganalysis, is the process of determining whether a particular instance of media contains hidden information and is thus a steganographic artifact.  A principle goal of steganography is resiliency to steganalysis such that if someone obtains the medium with the hidden information, they would not be able to detect its presence. The field of steganography has recently advanced due to the introduction of new media for use as stegonographic media and improved techniques for embedding information. Most modern approaches to the steganography problem focus on embedding information within images. As an aid in introducing the basic principles of steganography and steganalysis, we use the example of digital image files as the medium, or "cover image."  When a cover image has undergone steganographic processing, it is transformed into a corresponding "stego image" that, to a human viewer, is generally indistinguishable from the associated cover image.

Steganographic algorithms have become more sophisticated as they have evolved to become more resilient to steganalysis techniques. For example, a very simple method of steganography is to hide bits of information by replacing the least significant bit (LSb) of a cover image pixel with a bit comprising the hidden information. A more advanced steganographic method replaces the LSb of a cover image with that of a hidden message as in the simple approach, but only in those LSbs that correspond to least noticeable pixels of an image; thus, the appropriate least noticeable pixels of the cover image are first determined before LSb replacement occurs. Least noticeable pixels are generally those in an image that comprise a subregion with a relatively high-frequency content, or alternatively, significant color and brightness variations with regard to nearby pixels.  
The nature of human perception, and accordingly some machine-trained perception algorithms, is that a group of spatially-close pixels whose colors and brightness are slightly modified result in a region that is less discernable to slight differences rather than a region that is largely composed of pixels of the same color and brightness.  The crucial aspect of this particular approach to steganography is the identification of such regions in a cover image prior to embedding the hidden information.  
The S-UNIWARD algorithm \cite{suniward} relies upon a cost function whereby, for each cover image region being considered for embedding hidden information, an embedding cost is first estimated that indicates how likely the use of that particular area cause a steganaylsis approach to be successful in declaring the image was indeed a stego image. The theory of the S-UNIWARD algorithm is based upon the use of a distortion function called UNIWARD that is computed over an image region in the wavelet domain as a sum of relative changes of wavelet coefficients in the highest frequency undecimated subbands.  Therefore, the neighborhood of each pixel is assessed for the presence of discontinuities and indicates boundaries in the cover image wherein the embedding of hidden information should not occur. Research still persists in the area of steganography today, though many stegananlysis methods developed today use S-UNIWARD as a common baseline to compare error rates in detection of steganographic content. 

The S-UNIWARD algorithm requires a number of parameters to be specified to ensure the approach performs reliably. Since parameter values cannot be estimated from data, these values must often be set manually through general rules of thumb or trial and error.

Our approach uses a steganographic assistant convolutional neural network (SA-CNN) to modify the parameters of the S-UNIWARD algorithm to make it more effective and adaptive to the cover image thus increasing its resistance to discovery by steganalyzers. To achieve this improvement, we modified the S-UNIWARD algorithm to allow for specifying multipliers for certain parameters that are set within the algorithm. The parameters are chosen automatically with a goal to maximally confuse a pre-trained steganalysis network. These multipliers are predicted by using a convolutional neural network. For the continuous case, our SA-CNN attempts to predict the optimal multipliers directly, whereas for the discrete case, it selects a multiplier from a human-curated list. That is, we update the parameters using a loss function that maximizes the error rate of an algorithm designed to detect steganographic images. Our experimental results indicate that by processing a cover image with our SA-CNN, we can adjust appropriate parameters of the steganographic algorithm S-UNIWARD \cite{suniward} to make it more robust to exploitation by a common steganalyzer such as Yedroudj-Net \cite{yedroudj_2018}.

\section{Background and Related Work}
Steganography describes the process of hiding information in some medium, such as an image or audio file, so that if someone intercepts the medium with the hidden information, they would not be able to readily notice any information hidden in the image. Steganalysis is the process of determining whether the medium has hidden information. Our research focuses on steganography and steganalysis using the medium of images—in particular, our research uses established steganalysis tools to enhance existing steganography techniques. 

In steganography, an image without any hidden information is called a \textit{cover image}, and an image with hidden information is called a \textit{stego image}. Various methods have been established for classifying the difference between cover and stego-images using traditional image processing and information theoretic approaches \cite{thai_2014}, \cite{yedroudj_2018}. 

Currently, deep learning has arisen as a state-of-the-art way to distinguish between cover and stego-images. In deep learning, a neural network is optimized to reduce error in an objective function, typically using many neural network layers. One type of neural network commonly employed is a convolutional neural network (CNN), that is often used with images to recognize and react to various spatial patterns. These networks optimize the coefficients of convolutional filters via back-propagation to find custom extracted image features for a given classification task. Both steganographic and steganalytic approaches have employed CNNs. 

Because of their advantages, convolutional neural networks have become the de-facto standard in these communities due to their reliability and efficiency.

Initially, steganalysis was performed through hand-crafted algorithms. However, in 2009, Oplatková et al. \cite{oplatkova_2009} trained feedforward neural networks to detect steganographic content within images. Tan and Li \cite{tan_2014} began to address the issues by training three convolutional neural networks on steganalytic data. 
At the time, the Spatial Rich Model (SRM) was used as a binary classifier was the forefront of steganalyzers. However, Tan  believed that convolutional neural networks had the potential to replace SRM as there was some similarity in the architecture of the two. At the time, convolutional neural networks would either quickly converge to a poor, local solution or, when trying to avoid those local solutions, would diverge from any other solution. They were able to mitigate these issues by first training stacked convolutional auto-encoders and then initializing the convolutional neural network in the same configuration as the stacked convolutional auto-encoders. While they were not able to match the effectiveness of SRM, they were able to outperform other convolutional neural networks. 
Qian \textit{et al.} \cite{qian_2015} built upon this approach using applied CNNs to perform steganalysis, classifying cover and stego-images from one another. Following this line of research, several neural networks such as Ye-Net \cite{ye_2017}, Xu-Net \cite{xu_2016}, and Yedroudj-Net \cite{yedroudj_2018} have achieved error rates as low as 31.2\%, 27.2\%, and 22.8\% \cite{yedroudj_2018} respectively when trained against S-UNIWARD with an embedding rate of 0.4 bits per pixel (bpp).

More recently, Sharifzadeh \textit{et al.} derived a more statistically based steganographic algorithm, called adaptive batch size image merging (AdaBIM) \cite{sharifzadeh_2020}. They model the cover and stego messages as Gaussian variables, and maximize the error of three statistic-based steganalyzers. They also extend their solution into the realm of cost-based steganography by applying their approach on top of cost-based steganographic algorithms, including S-UNIWARD. However, their research is focused on batch steganography, in which information is hidden across multiple cover images, as opposed to our work. Additionally, they use a different kind of steganalyzer, using an ensemble classifier rather than a deep neural network.

\section{Method and Results}

\subsection{Dataset Collection}

To evaluate and inform the design of our model, we use the BOSSBase 1.01 \cite{bossbase} image database. This image database contains 10,000 grayscale images in the pgm image format. We used this database because it is freely available, its images are in a lossless format (and therefore compatible with S-UNIWARD), and it is the database used to train the steganalyzer we used in our research, Yedroudj-Net. As Yedroudj-Net was initially published using images of size 256x256, whereas the BOSSBase 1.01 images are originally at a size of 512x512, we followed the example of Yedroudj-Net and used bilinear resampling to scale each image to half its original width and height.

\subsection{Steganographic Method Manipulation}
For our steganographic embedding algorithm, we used the freely available S-UNIWARD algorithm. The S-UNIWARD algorithm and its variants are content-adaptive, effectively meaning that they hide information in the areas that they calculate will be least likely to reveal steganographic content. S-UNIWARD and its variants are widely used in research for steganography because they are advanced steganographic algorithms that embed dummy data at a specified rate and cause high error rates for steganalyzers, especially at lower embedding rates. For our research, we used an embedding rate of 0.4 bits per pixel (bpp). S-UNIWARD also requires three modifiable parameters. These parameters are usually set by users of S-UNIWARD for information hiding. However, we choose to control these parameters automatically using a SA-CNN to analyze the cover image. In this way, SA-CNN fine-tunes the parameters of S-UNIWARD according to the cover image of interest.

The three parameters of S-UNIWARD that we optimize are $\sigma$, $\epsilon$, and \textit{wet cost}. The $\sigma$ value determines to what degree the S-UNIWARD algorithm should value embedding data smoothly, at the cost of embedding data in more noticeable locations. The $\epsilon$ value is not strictly a parameter of the S-UNIWARD method. Rather, it is a constant set by the MATLAB implementation used for pre-processing logic of the S-UNIWARD algorithm. This value is used as a tolerance value for a value ``close enough'' to 0 or 1 to be rounded to the nearest whole number. We treat this value as a parameter by passing a value into the S-UNIWARD algorithm to multiply with $\epsilon$, using the resulting product in the S-UNIWARD algorithm. Wet cost is a value assigned to pixels that should not be modified. Practically, the value should be large enough so that the only pixels that are modified have a low probability of detection.

\subsection{Targeted Steganalyzer}
For our exemplary steganalyzer, we selected Yedroudj-Net \cite{yedroudj_2018}. In their research, Yedroudj \textit{et al.} showed improvements over other similar neural network steganalyzers, with error rates for S-UNIWARD reported as low as 36.7\% and 22.8\% for images encoded at 0.2 bpp and 0.4 bpp, respectively. Some of their improvements over previous techniques include more fully connected layers, more features per convolutional layer, and using a 30 filter bank for preprocessing. Yedroudj-Net also has an open-source implementation available allowing us to easily incorporate customizations. We used Yedroudj-Net as a steganalyzer over the BOSSBase 1.01 dataset, and verified our results by reproducing the results of Yedroudj-Net \cite{yedroudj_2018}.

\subsection{Convolutional Neural Network Design and Training}
\begin{figure}[ht]
  \centering
  \includegraphics[width=\linewidth]{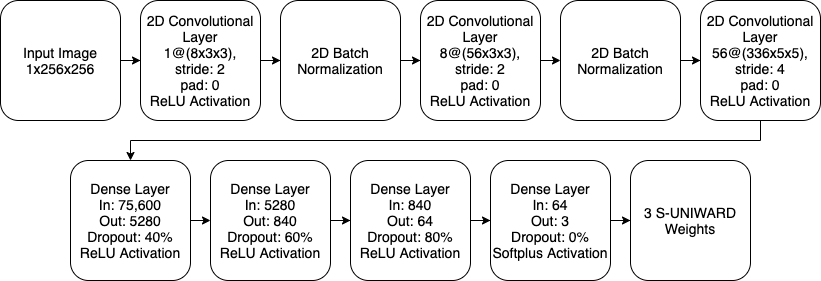}
  \caption{Convolutional Neural Network Structure}
  \Description{The structure of our Convolutional Neural Network}
  \label{fig:1 cnn structure}
\end{figure}

For our SA-CNN, we used the structure in Fig. \ref{fig:1 cnn structure}: 3 convolutional layers with batch normalization and strided down sampling, followed by fully connected layers with 40\%, 60\%, and 80\% dropout, respectively. We then used a softplus activation function to map the neural network output to a factor by which we could multiply the standard parameters of S-UNIWARD. The loss function we used is described below. We also used an exponential learning rate scheduler for our AdaM optimizer. We used a 90\%/10\% split for our train and validation set.

\begin{figure}[ht]
    \centering
    \includegraphics[width=8cm]{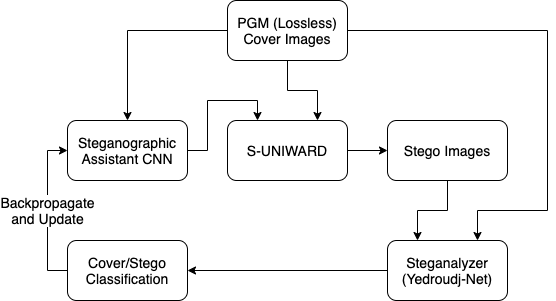}
    \caption{Neural Network Training Method}
    \Description{The training method for our Convolutional Neural Network}
    \label{fig:2 training diagram}
\end{figure}

Our training process followed the sequence shown in Fig. \ref{fig:2 training diagram}. Using our SA-CNN, we made an initial guess for fine-tuning the parameters of S-UNIWARD for each image in the dataset. We input those “tweaked” parameters with their corresponding cover images into S-UNIWARD to manufacture the stego-images. We then passed these cover-stego pairs as input to the Yedroudj-Net model and used its output, for each image, to determine how effective our SA-CNN performed against Yedroudj-Net. We then used these results to backpropagate and update our neural network until it had trained for 140 epochs (about seven days on our high-performance computing cluster, SMU Maneframe II). We then compared Yedroudj-Net’s error rate after training to the baseline reported error rate from \cite{yedroudj_2018} on the same dataset. A higher error rate on the Yedroudj-Net model means that our model’s input made S-UNIWARD more effective, as the steganographic embedding was more difficult to detect. Therefore, our objective function is to maximize the error rate of the Yedroudj-Net model.

\subsection{Experimental Results}
When training the SA-CNN, we tracked the error rate of both the training and validation set as a way to understand if the neural network was overfitting. During training, a higher error rate of the steganalyzer meant that the steganalyzer was fooled more often, meaning our SA-CNN increased the effectiveness of S-UNIWARD. Each time the validation set reached a new global high in the steganalyzer’s error rate, we would save the model. We tracked the training error rates of the steganalyzer throughout the training cycle. We then measured the validation accuracy by iterating through the entire validation set and tracking the accuracy of the classifications the steganalyzer made. In our evaluation, we use data from additional image databases to understand SA-CNN performance. 

\subsubsection{Similar Dataset}
For our initial evaluation, we ask: \textit{Is the increased error rate with our SA-CNN consistent when using similar images to the training dataset?} To answer this question, we manipulated the BOSSbase dataset, where instead of scaling down to half its original width and height, we cropped the image to the bottom-right quarter of the image such that images were similar but not identical to training data. We then ran the entirety of this dataset through S-UNIWARD and the familiar Yedroudj-Net model to calculate the baseline error rate, and then through SA-CNN, S-UNIWARD, and Yedroudj-Net for each model's confusion matrix and error rate. From the confusion matrix (Table~\ref{tab1}, SA-CNN \& Familiar Model columns), we observe that the new error rate, when the classifier thinks a cover image is a stego-image or not, is around 36.3\%, a substantial increase of 8.1 percentage points over the baseline reported error rate of 28.2\%. This means that the Yedroudj-Net model is being fooled more often, which would indicate that SA-CNN and S-UNIWARD combined are more effective for this dataset than S-UNIWARD for the original test set.

\begin{table*}[htp]
    \caption{Confusion matrix for two distinct Yedroudj-Net models on cropped BOSSbase dataset}
    \begin{center}
        \begin{tabular}{|c|c|c|c|c|c|c|}
            \hline
            & \multicolumn{2}{|c|}{\textbf{S-UNIWARD Only}} & \multicolumn{2}{|c|}{\textbf{SA-CNN \& Familiar Model}}&\multicolumn{2}{|c|}{\textbf{SA-CNN \& Unfamiliar Model}} \\
            \cline{2-7} 
            & \textbf{Cover Class} & \textbf{Stego Class}& \textbf{Cover Class} & \textbf{Stego Class} & \textbf{Cover Class} & \textbf{Stego Class} \\
            \hline
            Cover Image & 35.7\% & 14.3\% & 32.2\% & 17.8\% & 32.4\% & 17.6\% \\
            \hline
            Stego Image & 13.9\% & 36.1\% & 18.5\% & 31.5\% & 19.6\% & 30.4\% \\
            \hline
            Error Rate & \multicolumn{2}{|c|}{28.2\%} & \multicolumn{2}{|c|}{36.3\%} & \multicolumn{2}{|c|}{37.2\%} \\
            \hline
        \end{tabular}
        \label{tab1}
    \end{center}
\end{table*}

To further verify our results, we trained a second Yedroudj-Net model following the same steps as before but held constant the trained SA-CNN model. This second trial was meant to help understand if the SA-CNN was simply exploiting the specifics of one trained Yedroudj-Net model, or if it was actually improving the S-UNIWARD algorithm against multiple steganalyzers. From Table~\ref{tab1} (right), we can see that the error rates are similar, with an error rate of 37.2\% (14.4 percentage points over the baseline) compared to the original 36.3\%. Therefore, indicating that SA-CNN is not simply exploiting an artifact in a particular Yedroudj-Net model.

\begin{table*}[htp]
    \caption{Confusion matrix for ILSVRC dataset}
    \begin{center}
        \begin{tabular}{|c|c|c|c|c|}
            \hline
            & \multicolumn{2}{|c|}{\textbf{S-UNIWARD Only}}&\multicolumn{2}{|c|}{\textbf{S-UNIWARD + SA-CNN}} \\
            \cline{2-5} 
            & \textbf{Cover Class} & \textbf{Stego Class} & \textbf{Cover Class} & \textbf{Stego Class} \\
            \hline
            Cover Image & 25.0\% & 25.0\% & 23.8\% & 26.2\% \\
            \hline
            Stego Image & 21.0\% & 29.0\% & 22.9\% & 27.1\% \\
            \hline
            Error Rate & \multicolumn{2}{|c|}{46.0\%} & \multicolumn{2}{|c|}{49.1\%} \\
            \hline
        \end{tabular}
        \label{tab2}
    \end{center}
\end{table*}

\subsubsection{Dissimilar Dataset}
We now ask: \textit{Does our SA-CNN model generalize to datasets that are dissimilar to their training data?} To answer this, we randomly selected 10,000 images from the ImageNet ILSVRC 2010 dataset \cite{russakovsky2015imagenet} (none of these images appear in the BOSSBase database). We resized these images to the same 256x256 pixel size and converted the images from RGB JPEG images to the grayscale PGM images that Yedroudj-Net and the SA-CNN were trained with. We tested Yedroudj-Net with these images twice: once with S-UNIWARD, and once with our SA-CNN+S-UNIWARD. We do not retrain the Yedroudj-Net or SA-CNN models on this data; it is only used for evaluation. Moreover, we used the second BOSSBase Yedroudj-Net model that our SA-CNN had not trained against. From our results (Table~\ref{tab2}), we find that, in an unrelated set of images, our SA-CNN improves the performance of S-UNIWARD, from an error rate of 46.0\% to 49.1\%. This represents an increased error rate by 3.1 percentage points, as opposed to the 13.5 and 14.4 percentage points previously observed. While numerically smaller, this error rate increase is compelling, as the ideal performance for a steganographic algorithm is for the steganalyzer to have an error rate of 50.0\%, as at this point, the steganalyzer would not be any more useful for detecting a stego-image than a random guess. The poor performance of Yedroudj-Net model on this new data is an indication it is overfit to the BOSSBase 1.01 dataset, but we are still able to increase its performance near to ideal.

\begin{table*}[htp]
    \caption{Confusion matrix for differing S-UNIWARD parameters}
    \begin{center}
        \begin{tabular}{|c|c|c|c|c|}
            \hline
            & \multicolumn{2}{|c|}{\textbf{$\sigma$ + $\epsilon$ + Wet Cost}}&\multicolumn{2}{|c|}{\textbf{$\sigma$}} \\
            \cline{2-5} 
            & \textbf{Cover Class} & \textbf{Stego Class} & \textbf{Cover Class} & \textbf{Stego Class} \\
            \hline
            Cover Image & 32.4\% & 17.6\% & 42.8\% & 7.2\% \\
            \hline
            Stego Image & 19.6\% & 30.4\% & 8.3\% & 41.7\% \\
            \hline
            Error Rate & \multicolumn{2}{|c|}{37.2\%} & \multicolumn{2}{|c|}{15.5\%} \\
            \hline
        \end{tabular}
        \label{tab3}
    \end{center}
\end{table*}

\subsubsection{$\sigma$ vs. $\sigma$ + $\epsilon$ + Wet Cost}
Next, we ask: \textit{Do the $\epsilon$ and wet cost values set by our SA-CNN model influence the effectiveness of our model?} To answer this, we only allow our SA-CNN to modify the $\sigma$ value of S-UNIWARD for the embedding process, leaving the $\epsilon$ and wet cost values at their default value. For the steganalyzer, we use the second Yedroudj-Net model, which our SA-CNN model did not interact with during training. For the dataset, we used the cropped BOSSBase dataset. We found that modifying the $\epsilon$ and wet cost values did have a noticeable impact on the effectiveness of our SA-CNN (see Table~\ref{tab3}). We believe this is likely due to the SA-CNN using the $\epsilon$ and wet cost values to force S-UNIWARD to embed a small amount of the payload into areas that are calculated to be incredibly costly, not embedding enough information to be reliably detected, but enough to reduce the amount of payload embedded into traditional embedding areas, making the payload less discernable.

\begin{figure}[ht]
  \centering
  \includegraphics[width=\linewidth]{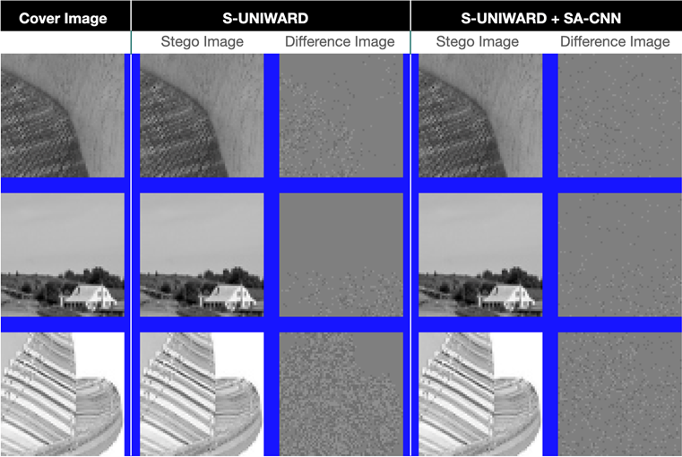}
  \caption{Example image results from our SA-CNN method}
  \Description{Example image results from our SA-CNN method}
  \label{fig:3 suniward comparisons}
\end{figure}

To investigate this hypothesis, we provide visualizations of some of the images generated with our SA-CNN+S-UNIWARD. Fig. \ref{fig:3 suniward comparisons} shows cover images from ImageNet with their respective stego-images from S-UNIWARD and SA-CNN+S-UNIWARD, as well as images with the changes amplified by a factor of eight. The cover image for each set is in the first column, followed by the S-UNIWARD images and their amplified changes in the second and third columns respectively, and finally the SA-CNN+S-UNIWARD images and their amplified changes in the fourth and fifth columns.

In general, we find that the stego-images generated by the SA-CNN+S-UNIWARD embedded data more smoothly than S-UNIWARD alone. The output of the SA-CNN supports this, as it consistently recommended higher $\sigma$ values than the default value. While more information can be hidden in higher-frequency areas without detection, we believe that the default S-UNIWARD is embedding too much data in the high-frequency areas of the cover images. Steganalyzers are then more likely to detect this excess data and correctly classify the image as a stego-image. The third row may appear to imply that the SA-CNN encourages S-UNIWARD to embed information in areas that are not typically selected for information hiding. This is an interesting result as it implies that, because steganalyzers tend to overlook information hidden in smooth areas, that hiding information in these areas is an efficient method for fooling existing steganalyzers. However, due to the results from Table~\ref{tab1}, we believe that the Yedroudj-Net models we trained were less receptive of changes in saturated areas. We believe that, when trained against a steganalyzer more alert to payloads embedded in highly saturated areas, the SA-CNN will adapt to the steganalyzer and emphasize embedding the payload in higher frequency areas.

\subsubsection{Discrete Output Space}
Finally, we ask: \textit{can we achieve comparable results with a discrete output space?} The advantages of such an approach are as follows. Firstly, we can pre-compute every possible output in the discrete space, saving time during the training of the neural network. Secondly, the simplification of the neural network's task from a regression task to a classification task could allow the neural network to converge in fewer epochs. However, there are a few drawbacks to this approach. First, every possible output needs to be generated before training the neural network, and while this is an easily parallelizable task, it does require a significant storage space, especially for the lossless images used by the S-UNIWARD algorithm. However, pre-computing all possible outputs is not required if the required storage space is so large that it outweighs the benefit gained from the decreased training time. Additionally, the neural network trained for the discrete output space may not achieve results as strong as the one trained for the continuous output space, as the optimal output for the neural network may not be represented in the discrete space.

\begin{figure}[ht]
    \centering
    \includegraphics[width=8cm]{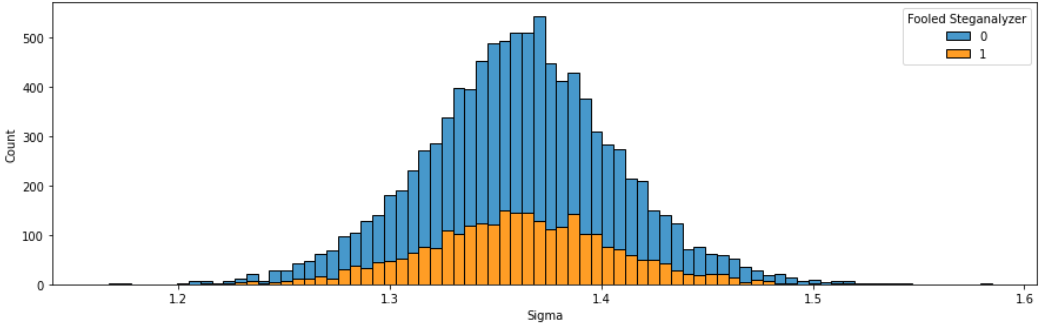}
    \caption{Continuous model $\sigma$ distribution}
    \Description{The distribution of the $\sigma$ values for the continuous model}
    \label{fig:4 continuous sigma}
\end{figure}

\begin{figure}[ht]
    \centering
    \includegraphics[width=8cm]{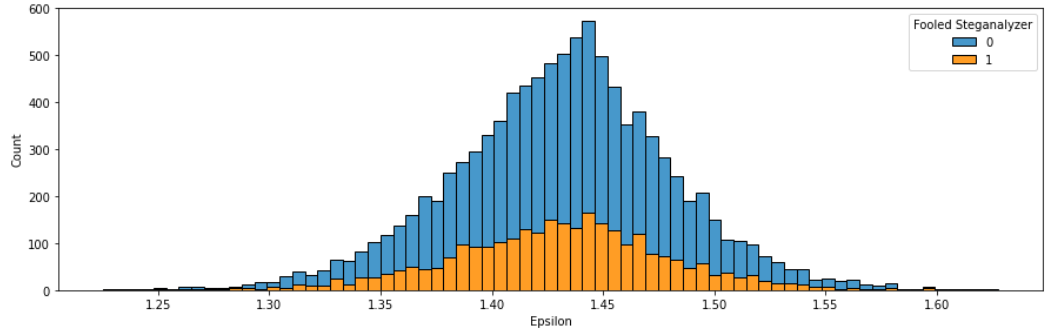}
    \caption{Continuous model $\epsilon$ distribution}
    \Description{The distribution of the $\epsilon$ values for the continuous model}
    \label{fig:5 continuous epsilon}
\end{figure}

\begin{figure}[ht]
    \centering
    \includegraphics[width=8cm]{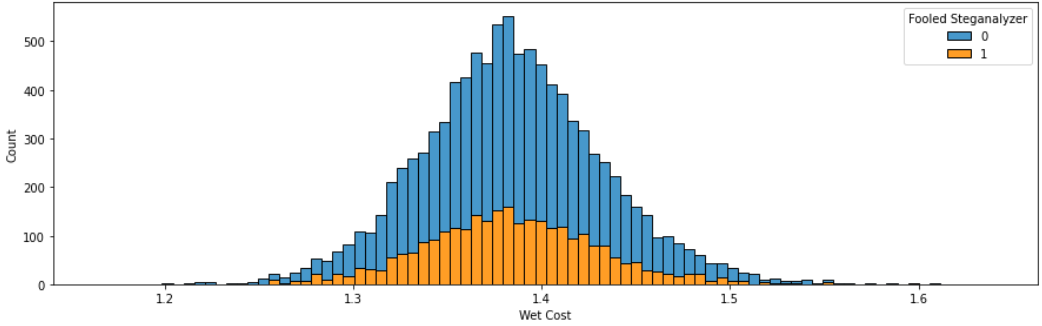}
    \caption{Continuous model wet cost distribution}
    \Description{The distribution of the wet cost values for the continuous model}
    \label{fig:6 continuous wet cost}
\end{figure}

\begin{table*}[htp]
    \caption{Comparison of continuous and discrete space outputs}
    \begin{center}
        \begin{tabular}{|c|c|c|c|}
            \hline
            & \textbf{Storage Space} & \textbf{Training Time} & \textbf{Validation Error Rate} \\
            \hline
            Continuous & 704 MB & 3796s $\pm$ 20s & 37.2\% \\
            \hline
            Stego Image & 2.8 TB & 723s $\pm$ 130s & 18.1\% \\
            \hline
        \end{tabular}
        \label{tab4}
    \end{center}
\end{table*}

To determine which possible outputs our discrete space model ought to be able to select from, we examined the distribution of outputs for our continuous space model. The distribution of each output type, $\sigma$, $\epsilon$, and wet cost, all formed a normal distribution, all centered around a value of 1.4 times the normal value, ranging between 1.2 times and 1.6 times the normal value. Using this information, we selected and fine-tuned a range of values between 1.3 and 1.5 for the discrete space, eventually arriving at the following distribution of 13 values: 1.3, 1.325, 1.35, 1.3625, 1.375, 1.3875, 1.4, 1.4125, 1.425, 1.4375, 1.45, 1.475, 1.5. As this same distribution is used for the $\sigma$, $\epsilon$, and wet cost values, this gives us an output discrete space of size 13 $\times$ 13 $\times$ 13. Below is several histograms of the distributions for each output from the continuous model, upon which we based the discrete model's distribution.

We evaluated our discrete space output model on the BOSSBase dataset using three metrics: storage space, training time per epoch, and validation accuracy. We compared each of these metrics to the continuous space equivalent.

First, we can see that the space required for the continuous space model is on the order of hundreds of megabytes, while the space required for the discrete space model is on the order of terabytes. However, the training time has decreased by over 5 times, which means that a continuous model that would ordinarily take half a week to train would train in around 16 hours as a discrete model.

We found that, over several trials, we were unable to get our discrete model to converge with a validation error rate better than that of the base S-UNIWARD implementation. While the continuous model boosted the YedroudjNet model's error rate by 14.4 percentage points, the discrete model reduced the YedroudjNet model's error rate by 4.1 percentage points, which is the opposite of our goal. To address this, we tried reducing the complexity of the model by removing some convolutional and fully connected layers. We tried varying the size and distribution of the discrete space from our original 7 $\times$ 7 $\times$ 7 to our final 13 $\times$ 13 $\times$ 13. We decided not to progress to distributions of size 17 $\times$ 17 $\times$ 17 or 26 $\times$ 26 $\times$ 26 as we estimated that pre-processing those discrete spaces would use roughly 6.3 TB and 22.4 TB of storage space respectively, which we deemed infeasible. We also tried modifying the shape of the output distribution, experimenting with both a uniform distribution and the non-uniform distribution at which we ultimately arrived.

We believe that the discrete space model's failure is due to one or several of the following factors. First, the sparseness of the discrete space may significantly reduce the number of images that have a valid combination of $\sigma$, $\epsilon$, and wet cost values that fool the steganalyzer, lowering the best possible error rate due to a lack of sufficient options. Second, the sparseness of the discrete space may make it too difficult to make fine adjustments to the model, especially for a task as sensitive as steganography. Finally, the discrete space that we chose might not have represented the optimal output distribution well enough for the model to fool the steganalyzer reliably. We do not think that the model was too complex to converge to the optimal solution, as we tested models ranging from 4 layers to 7 layers and all models showed similar results. We also do not think the problem was too complex for the model, as we used a nearly identical model architecture for both the discrete and continuous cases, modifying only the output layer for the discrete model. The difficulty of the discrete problem ought to be comparable, if not easier, than that of the continuous problem, as the range of possible output values is reduced from infinite to 13 $\times$ 13 $\times$ 13.

\section{Conclusion and Future Work}
Our results show that previously established steganographic algorithms can be improved by using a neural network to modify parameters of previously established methods. Specifically, we use S-UNIWARD as an example, increasing its error rate against a steganalyzer in various datasets. In particular, the results are more interpretable because they build upon an established method. Several areas for future research are still possible, including:

\begin{itemize}
    \item Training an SA-CNN against multiple steganalyzers at the same time.
    \item Expanding the current model to work with other steganographic methods, using generalizable outputs across different steganographic methods.
    \item Studying the impact of different embedding rates with the SA-CNN outputs for S-UNIWARD parameters.
    \item Using the adversarial training method for generative adversarial networks to train the SA-CNN and Yedroudj-Net against each other.
\end{itemize}
\typeout{}
\bibliographystyle{IEEEtran}
\bibliography{refs.bib}
\end{document}